\documentclass[final]{scri}
\usepackage{graphicx}

\begin{document}
\title{
Analytic Approximations for the Velocity of Field-Driven Ising Interfaces
}

\author{
Per Arne Rikvold$^1$ and M.~Kolesik$^{2,3}$
}

\date{\today}

\maketitle
\footnotetext[1]{
Center for Materials Research and Technology, 
School of Computational Science and Information Technology, and 
Department of Physics,
Florida State University, Tallahassee, FL 32306-4350.\\
E-mail: rikvold@csit.fsu.edu
}
\footnotetext[2]{
Institute of Physics, Slovak Academy of Sciences,
Bratislava, Slovak Republic
} 
\footnotetext[3]{
Department of Mathematics, University of Arizona,
Tucson, AZ 85721.\\
E-mail: kolesik@acms.arizona.edu 
}

\begin{abstract}
We present analytic approximations for the field, temperature, 
and orientation dependences 
of the interface velocity in a two-dimensional kinetic Ising model in a 
nonzero field. The model, which has nonconserved order parameter, 
is useful for ferromagnets, ferroelectrics, and other systems undergoing 
order-disorder phase transformations driven by a bulk free-energy difference. 
The Solid-on-Solid (SOS) approximation for the microscopic 
surface structure is used to estimate mean spin-class populations, from 
which the mean interface velocity can be obtained for any specific 
single-spin-flip dynamic. This linear-response 
approximation remains accurate for higher 
temperatures than the single-step and polynuclear growth models, 
while it reduces to these in the appropriate low-temperature limits. 
The equilibrium SOS approximation is generalized by 
mean-field arguments to obtain field-dependent 
spin-class populations for moving interfaces, and thereby a nonlinear-response 
approximation for the velocity. 
The analytic results for the interface velocity and the spin-class 
populations are compared with Monte Carlo 
simulations. Excellent agreement is found in a wide range of field, 
temperature, and interface orientation. 

\end{abstract}


\vspace{2.5truecm}
\noindent
{\Large {\it J.\ Stat.\ Phys.\/} {\bf 100}, in press. \hfill Cond-mat/9909188}

\clearpage

\section{Introduction}
\label{sec:INTRO}

The appearance of the world around us through sight and touch is 
largely determined by interfaces between phases with different optical and 
mechanical properties, and the materials properties 
of multiphase media are strongly influenced by the geometry of 
interfaces that separate their constituent materials \cite{STELL87,TORQ98}. 
Since, in general, the morphology of an
interface is determined by the growth process by which it is formed,
the large-scale structures of growing interfaces have inspired an
enormous amount of work in the last few decades \cite{BARA95,MEAK98}. 

In comparison to the vigorous interest in large-scale structure, 
much less attention has been paid to interfacial
structure on a microscopic scale. This is somewhat surprising
since the microscopic structure 
limits the interfacial propagation velocity under an external driving
force, such as the applied field for a magnetic or dielectric domain wall 
or the supersaturation or supercooling for a crystal surface.
It is also important for properties such as chemical reactivity and 
catalytic activity. 

In this paper we consider how the microscopic interface structure 
determines the growth velocity of a simple model surface in a 
system with nonconserved order parameter: 
the interface between domains of positive and negative magnetization in a
square-lattice kinetic Ising ferromagnet with nearest-neighbor interactions, 
which is driven by a field favoring 
one of the two spin orientations \cite{DEVI92,SPOH93,SHNE99A}. 
This model is applicable to the kinetics of phase transformation in many 
magnetic and ferroelectric systems and other order-disorder 
transitions whose kinetics are not inhibited by coupling to a conserved 
field. It belongs to the
dynamic universality class of the Kardar-Parisi-Zhang (KPZ)
model \cite{KARD86}, and the velocity of a macroscopically plane interface is 
expected to be linear in an asymptotically weak field, as is also predicted 
by the Lifshitz-Allen-Cahn theory \cite{LIFS62,ALLE79}. 
However, neither theory gives the explicit field dependence, which should 
contain both the average interface orientation and the specific dynamic. 
Here we derive analytic, approximate expressions for the mean velocity 
as a function of field, temperature, and interface orientation. 
Our approach is based on the concept of {\it spin classes\/} used in
rejection-free Monte Carlo (MC) algorithms \cite{BORT75,NOVO95A,NOVO95},
together with the Burton-Cabrera-Frank Solid-on-Solid
(SOS) approximation for the structure of a stationary interface
\cite{BURT51,KRUG95}. While the theory should become exact for asymptotically 
small temperatures and fields, our main purpose is to explore its 
applicability outside this limited regime. 

The remainder of this paper is organized as follows. 
In Sec.~\ref{sec:MODEL} we introduce the kinetic Ising model and 
the concept of spin classes.
In Sec.~\ref{sec:SOS} we summarize relevant aspects of the 
SOS approximation for the structure 
of a flat, equilibrium Ising interface between two bulk phases of opposite 
magnetization, which we use to obtain analytic approximations 
for the mean spin-class populations in zero field. These provide a 
linear-response approximation for the velocity of a driven interface. 
In Sec.~\ref{sec:NLR} we develop an extension of the SOS approximation to 
obtain field-dependent spin-class populations for flat, moving interfaces 
as well. This leads to a nonlinear-response approximation for the 
velocity. In Sec.~\ref{sec:MC} 
the theoretical results for the interface velocity and spin-class
populations from Secs.~\ref{sec:SOS} and~\ref{sec:NLR} 
are compared with MC simulations. 
The nonlinear-response approximation gives remarkable agreement 
with the simulations in a wide range of field, temperature, and interface 
orientation. Section \ref{sec:DISC} contains a discussion, conclusions,
and some suggestions for future work.

\section{Model and Dynamics}
\label{sec:MODEL}

The anisotropic square-lattice  Ising ferromagnet with nearest-neighbor 
interactions is defined by the Hamiltonian 
\begin{equation}
{\cal H} = -\sum_{x,y} s_{x,y} \left( J_x s_{x+1,y} + J_y s_{x,y+1} 
+ H \right) 
\;. 
\label{eq:ham}
\end{equation}
Here $s_{x,y} = \pm 1$, $\sum_{x,y}$ runs over all lattice sites, 
and $H$ is the applied field. The lattice constant is taken as our 
unit of length. An interface is introduced by fixing 
$s_{x,y}=+1$ and $-$1 for large negative and positive $y$, respectively.  
{}For concreteness we assume that $H \ge 0$, such 
that the interface moves in the positive $y$ direction under an applied field. 
The implementation of these boundary conditions in our MC simulations is 
discussed in Sec.~\ref{sec:MCs}. 

Approach to equilibrium is ensured by a single-spin-flip 
(nonconservative) dynamic which satisfies detailed balance, such as the 
Metropolis or Glauber algorithms \cite{KAWA72}. 
Any such algorithm is defined by a transition probability, 
$W[s_{x,y} \rightarrow -s_{x,y}] = W[\beta \Delta E]$, where 
$\beta$ is the inverse of the temperature $T$ (we use units in which 
Boltzmann's constant is unity), and $\Delta E$ is the energy change that 
would occur if the proposed spin flip were accepted. 
Since there are only a finite 
number of different values of $\Delta E$, the spins can be 
divided into classes \cite{SPOH93,BORT75}, 
labeled by the spin value $s$ and the number 
of broken bonds between the spin and its nearest neighbors in the $x$- and 
$y$-direction, $j$ and $k$, respectively. 
The spin classes, denoted $jks$ with $j,k \in \{0,1,2\}$, 
are listed in Table~1 
together with the corresponding energies, $E(jks)$, 
and energy changes, $\Delta E(jks)$. For the anisotropic model defined 
by Eq.~(\ref{eq:ham}) there are 18 classes \cite{NOVO95A}. 
In the isotropic case, 
$J_x=J_y$, this reduces to 10 classes distinguished by $s$ and the 
total number of broken bonds, $j$+$k$ \cite{BORT75}. 

{}For concreteness and comparison with 
numerical simulations we here choose the discrete-time 
Glauber dynamic, defined by the transition probability 
\begin{equation}
W_{\rm G}[s_{x,y} \rightarrow -s_{x,y}] 
= \frac{e^{ - \beta \Delta E }}{1 + e^{- \beta \Delta E }}
\;.
\label{eq:glau}
\end{equation}
The Glauber dynamic is mathematically convenient in that the transition 
probability is a continuously differentiable function of $\Delta E$. 
In its continuum-time version it 
has also been shown to correspond to a quantum-mechanical $S$=1/2 system 
weakly coupled to a large thermal fermion bath \cite{MART77}. 
However, the spin-class populations can be used to estimate 
propagation velocities with any single-spin-flip dynamic that satisfies 
detailed balance. Time is measured in units of MC steps per spin (MCSS). 

To prevent nucleation of droplets of the stable phase in front of the moving 
interface \cite{DEVI92}, 
we modify the Glauber dynamic by setting the transition rate for 
any spin which is parallel to all its neighbors (i.e., class $00s$) equal 
to zero \cite{SHNE99A,SHNE98,RAMO99}.  This suppresses  
thermal fluctuations in both the bulk phases, 
while the local interface structure is reasonably 
preserved. For moderate fields the interface 
velocity with this modified Glauber dynamic is only slightly less than that 
obtained with the full Glauber dynamic \cite{SHNE98,RAMO99}. 
In the strong-field regime, 
where the size of a critical droplet of the equilibrium phase 
is reduced to on the order of the lattice constant, the kinetic Ising 
interface loses its integrity. A conservative analytic approximation for 
this crossover field (often called ``the mean-field 
spinodal'' \cite{TOMI92A,RIKV94A}) is 
(for isotropic interactions) \cite{RICH94} 
$H_{\rm MFSP}(T) \approx \sigma(T)/m_{\rm eq}(T)$, 
where $\sigma(T)$ and 
$m_{\rm eq}(T)$ are the equilibrium surface tension in the $x$-direction and 
the equilibrium magnetization, respectively.  In the strong-field 
regime the SOS approximation 
and the dynamic used here should be considered as a 
nonequilibrium cluster growth model in its own right. 
{}For $|H| < H_{\rm MFSP}(T)$ they constitute a good 
approximation for the kinetic Ising model with the full Glauber dynamic 
\cite{SHNE98,RAMO99}.

\section{The SOS Approximation}
\label{sec:SOS}

The separation of spins into classes forms the basis of several  
rejection-free MC algorithms 
\cite{BORT75,NOVO95A,NOVO95,KOLE98A,KOLE98B,NOVO99}. 
In such algorithms the spin-class populations, $n(jks)$, 
are continually monitored throughout the simulation. 
Given this information and 
the transition probabilities of the particular dynamic 
used, one can then calculate the time increments between MC updates. 
Here we instead obtain analytic approximations for the  
mean spin-class populations for a driven 
interface moving at a constant velocity, based on the Burton-Cabrera-Frank  
SOS model of the equilibrium interface 
\cite{BURT51}. These populations are then used together with the 
transition probabilities to obtain the mean interface velocity. 

\begin{figure}[ht] 
\begin{center}
\includegraphics[angle=270,width=5in]{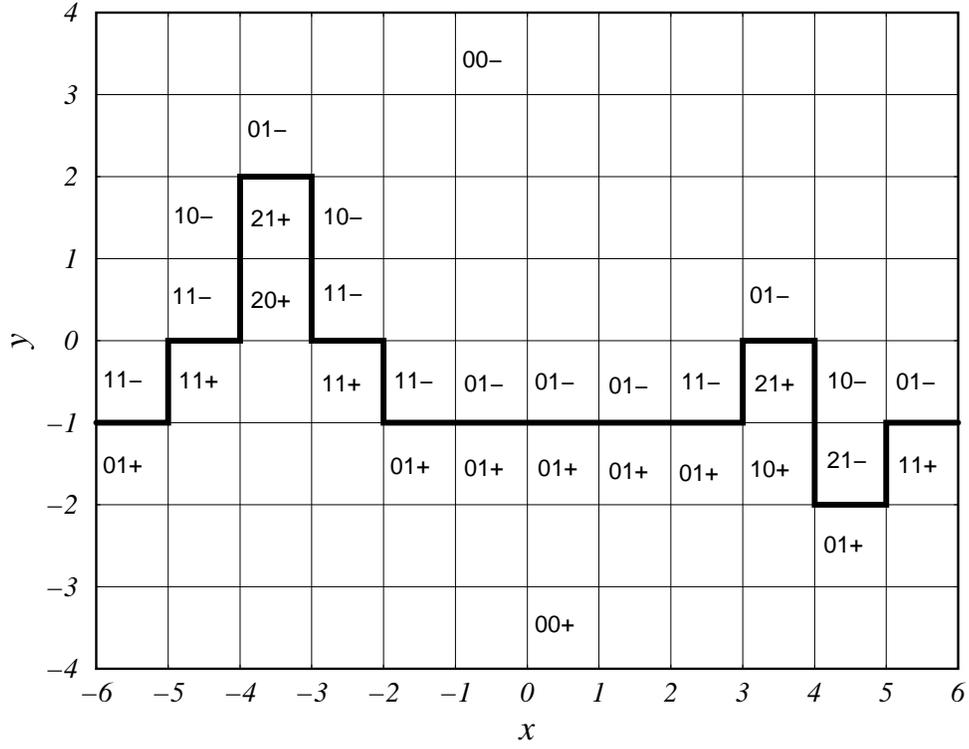}
\end{center}
\caption[]{
A short segment of a zero-field equilibrium
SOS interface between a positively magnetized phase for 
$y < 0$ and a negative phase for $y>0$. The independent step heights, 
$\delta(x)$, are drawn from the pdf given by Eqs.~(\ref{eq:step_pdf}) 
and~(\ref{eq:Z}) with $T = 0.6 T_c$ and $\phi = 0$. 
Interface sites representative of 
the different spin classes compatible with the SOS approximation are marked 
with the notation $jks$ explained in Sec.~\ref{sec:MODEL}. 
Sites in the uniform bulk phases are $00-$ and $00+$. 
}
\label{fig:SOS}
\end{figure}

The SOS approximation describes the interface as a single-valued function 
$y(x)$. For the square lattice considered here, the interface is a series 
of integer-valued steps of height $\delta(x)$ parallel to the $y$-axis, 
as shown in Fig.~\ref{fig:SOS}. 
The heights of the individual 
steps are assumed to be statistically independent and 
identically distributed. The probability density function (pdf) 
is given by the interaction energy corresponding to the $|\delta(x)|$ broken 
$J_x$-bonds between spins in the columns centered 
at $(x-1/2)$ and  $(x+1/2)$ as  
\begin{equation}
p_x[\delta(x)] = Z_x^{-1} X^{|\delta(x)|}
\ e^{ \gamma(\phi) \delta(x) } \;, 
\label{eq:step_pdf}
\end{equation}
where we have introduced the shorthand $X = e^{- 2 \beta J_x}$. 
Here $\gamma(\phi)$ is a Lagrange multiplier which maintains the mean step 
height at an $x$-independent value, $\langle \delta(x) \rangle = \tan \phi$, 
where $\phi$ is the overall angle between the interface and the $x$-axis. 
The partition function is 
\begin{equation}
Z_x 
=
\sum_{\delta = -\infty}^{+\infty} 
X^{|\delta|} e^{ \gamma(\phi) \delta } \nonumber\\
= 
\frac{1-X^2}{1 - 2 X \cosh \gamma(\phi) + X^2} 
\;. 
\label{eq:Z}
\end{equation}
Using $Z_x$ as a moment generating function for $\delta(x)$, it 
is straightforward to obtain the explicit expression
\begin{equation}
e^{\gamma (\phi)} 
= 
\frac{ \left(1+X^2 \right)\tan \phi + R}{2 X \left( 1 + \tan \phi \right)} 
\;,
\label{eq:chgam}
\end{equation}
where $R = \left[ \left( 1 - X^2 \right)^2 \tan^2 \phi + 4 X^2 \right]^{1/2}$.
Combining Eqs.~(\ref{eq:Z}) and (\ref{eq:chgam}) 
one obtains $Z_x$ explicitly as a function of $\phi$: 
\begin{equation}
Z_x(\phi) = \frac{ \left( 1 - X^2 \right) \left( 1 - \tan^2 \phi \right)}
{1 + X^2 - R}
\;.
\label{eq:Zphi}
\end{equation}
{}For $\phi=0$ this simplifies to $Z_x(0) = (1+X)/(1-X)$. 
The SOS approximation ignores overhangs and bubbles. 
It is therefore rather remarkable that the surface tension in 
this approximation, calculated as 
$\sigma_{\rm SOS} = 
|\cos \phi | \left[ 2 J_y - T \ln Z_x(\phi) + 
T \gamma(\phi) \tan \phi \right]$, 
yields the exact result  
for $\phi=0$ \cite{TEMP52} and an excellent approximation for 
$|\phi| \le \pi/4$ \cite{AVRO82}. For larger $|\phi|$ it is more reasonable 
to use an SOS approximation with steps parallel to the $x$-axis. 

While Eq.~(\ref{eq:Zphi}) is equivalent to Eq.~(72) of Ref.~\cite{BURT51}, 
the implicit form given by Eqs.~(\ref{eq:Z}) and (\ref{eq:chgam}) 
is more convenient for our purpose of obtaining mean spin-class 
populations, $\langle n(jks) \rangle$. 
The spin classes compatible with this approximation are illustrated in 
Fig.~\ref{fig:SOS}. The mean populations are all obtained from the 
joint pdf for $\delta(x)$ and $\delta(x$+1). Since the 
individual step heights are statistically independent,
this is the product $p_x[\delta(x)] \cdot p_{x+1}[\delta(x$+1)]. 
The symmetry of $p_x[\delta(x)]$ under the transformation 
$(x,\phi,\delta) \rightarrow (-x,-\phi,-\delta)$ ensures that 
$\langle n(jk-) \rangle = \langle n(jk+) \rangle$ for all $j$ and $k$. 
(On the right-hand sides of Eqs.~(\ref{eq:01s}) -- (\ref{eq:pnn}) below, 
we have chosen $s = -1$ with the interface oriented as shown in 
Fig.~\ref{fig:SOS} for concreteness.) 

The SOS picture implies that there is exactly one broken 
$J_y$ bond per unit length in the $x$-direction, so that 
$\langle n(01s) \rangle + 
\langle n(11s) \rangle + \langle n(21s) \rangle = 1$. 
The calculations of the individual populations 
are straightforward but somewhat tedious, especially for nonzero $\phi$.   
In Eqs.~(\ref{eq:01s})--(\ref{eq:21s}) below we 
therefore just give the starting point of the calculation
for each class in terms of 
$p_x[\delta(x)]$ and the cumulative probabilities
$P[\delta(x) \le n] = \sum_{\delta = - \infty}^n p_x(\delta)$ 
and $P[\delta(x) \ge n] = 1 - P[\delta(x) \le (n-1)]$. The 
final results are listed in Table~2, 
both for general $\phi$ and for $\phi=0$. 
\begin{eqnarray}
\langle n(01s) \rangle 
&=& 
P[\delta(x) \ge 0] \cdot P[\delta(x+1) \le 0] 
\label{eq:01s}
\\
\langle n(11s) \rangle 
&=& 
P[\delta(x) \le -1] \cdot P[\delta(x+1) \le 0] 
+
P[\delta(x) \ge 0] \cdot P[\delta(x+1) \ge 1] 
\label{eq:11s}
\\
\langle n(21s) \rangle 
&=& 
P[\delta(x) \le -1] \cdot P[\delta(x+1) \ge 1] 
\label{eq:21s}
\end{eqnarray}
Flipping a spin in either of these 
classes (marked $\ast$ in Table~2) 
preserves the SOS configuration. 

To obtain the mean populations for classes of spins that are connected to the
interface only through one (10$s$) or two (20$s$) broken $J_x$ bonds is 
more tedious. We found it most convenient first to calculate 
the joint pdf for $n = n(10s) + n(20s)$ and  $n(20s)$:
\begin{equation}
p[n,n(20s)] 
= 
\left \{ 
\begin{array}{ll}
P[\delta(x) \ge -1] \cdot P[\delta(x+1) \le 1] 
  & \mbox{for $n=n(20s)=0$} \\
\\
p_x[-(n+1)] \cdot P[\delta(x+1) \le 1] 
\\
+
P[\delta(x) \ge -1] \cdot p_{x+1}[n+1] 
  & \mbox{for $n \ge 1$, $n(20s)=0$} \\
\\
p_x[-(n+1)] \cdot p_{x+1}[n(20s)+1] 
\\
+
p_x[-(n(20s)+1)] \cdot p_{x+1}[n+1] 
  & \mbox{for $n > 1$, $1 \le n(20s) < n$} \\
\\
p_x[-(n+1)] \cdot p_{x+1}[n+1] 
  & \mbox{for $n(20s) = n \ge 1$}  
\end{array} 
\right. 
\label{eq:pnn}
\end{equation}
The resulting expressions for 
$\langle n(10s) \rangle$ and $\langle n(20s) \rangle$ are 
marked $\dag$ in Table~2. 
Flipping a spin in one of these classes results in the breaking of two 
$J_y$-bonds, and thus in the creation of an overhang or a bubble. 
The classes, 02$s$, 12$s$, and 22$s$, are not populated in 
the SOS approximation, while 00$s$ represents bulk sites which have zero 
transition rates with this dynamic. 

Each column of the interface advances by one lattice constant 
in the $y$-direction whenever a spin  
flips from $-$1 to +1, regardless of its 
$y$-coordinate. Conversely, the interface  
recedes by one lattice constant whenever a spin 
flips from +1 to $-$1. The energy changes corresponding to a flip are 
given in the third column in Table~1. 
Since the spin-class populations on both sides of the 
interface are equal in this approximation, the contribution 
from sites in the classes $jk-$ and $jk+$ 
to the mean velocity  in the $y$-direction 
is the difference between the transition 
probabilities for spin flips leading to advance and recession:
\begin{equation}
\langle v_y(jk) \rangle 
= 
W \left[ \beta \Delta E(jk-)  \right]
-
W \left[ \beta \Delta E(jk+)  \right] 
 \;. 
\label{eq_generalv}
\end{equation}
The results corresponding to 
the Glauber transition probabilities from Eq.~(\ref{eq:glau}) 
are given in the last column in Table~2.  
(It is of course trivial to generalize to 
$\langle n(jk-) \rangle \neq \langle n(jk+) \rangle$, but this will not 
be needed here.) 
The mean propagation velocity perpendicular to the interface becomes  
\begin{equation}
\langle v_\perp (T,H,\phi) \rangle 
= 
|\cos \phi | \sum_{j,k} \langle n(jks) \rangle \langle v_y (jk) \rangle 
\;. 
\label{eq:totalv}
\end{equation}
Including the SOS-violating moves, $jk \in \{10, 20\}$, in principle allows  
the propagation velocity to exceed unity as it becomes possible to flip
several spins along a high vertical step. However, 
such large velocities are only
observed for the strongest fields investigated here. 
Restricting the sum to only $jk \in \{01, 11, 21\}$, on the other hand,  
yields an approximation which excludes SOS-violating transitions and
would limit the velocity to below unity. 

While the general result for the velocity is rather cumbersome 
if written out in detail, the special case of $\phi=0$ leads to a relatively 
compact formula: 
\begin{eqnarray}
\langle v_\perp (T,H,0) \rangle 
&=&
\frac{\tanh (\beta H)}{(1+X)^2} 
\left\{
2 X 
+ 
\frac{1+X^2}
{1 + \left[\frac{\sinh (2 \beta J_x)}{\cosh (\beta H)}\right]^2} \right.
\nonumber\\
& & 
+
\left. \frac{X^2}{1-X^2}
\left[
\frac{X^2} 
{1 + \left[\frac{\sinh (2 \beta (J_y-J_x))}{\cosh (\beta H)}\right]^2} 
+ 
\frac{2 (1+2X)}
{1 + \left[\frac{\sinh (2 \beta J_y)}{\cosh (\beta H)}\right]^2} 
\right]
\right\} 
\;.  
\label{eq:totalv0}
\end{eqnarray}
Here the first line corresponds to transitions that preserve the SOS 
structure of the interface, while the second line corresponds to 
transitions that create overhangs or bubbles. 
Comparison with simulation data indicate that excluding the 
SOS-violating transitions 
leads to significant underestimation of the propagation velocity, 
even for quite moderate fields. This effect is shown in Fig.~\ref{fig:veloH} 
and discussed in more detail in Sec.~\ref{sec:MCi}. 

Equation~(\ref{eq:totalv0}) with $J_x=J_y$
was presented without detailed derivation in 
Ref.~\cite{RAMO99}. In that work the average 
interface velocity in a kinetic Ising model undergoing a field-driven phase 
transformation was estimated directly from the time evolution of the 
magnetization after a sudden field reversal. 
This estimate was found to be consistent with Eq.~(\ref{eq:totalv0}).  

As $T \rightarrow 0$, $X \rightarrow 0$. In this limit, 
the dominant term in Eq.~(\ref{eq:totalv}) is 
the one proportional to $\langle n(11s) \rangle$, 
which is simply the density of kinks on the surface. 
Combining the appropriate entries 
in Table~2 with Eq.~(\ref{eq:chgam}) 
and ignoring $X$ everywhere except where $X^2$ occurs in an
additive combination with $\tan^2 \phi$, we 
get the angle-dependent interface velocity for very low $T$: 
\begin{eqnarray} 
{\langle v_\perp (T \rightarrow 0, H, \phi) \rangle}
&=&
\cos \phi 
\frac{\sqrt{\tan^2 \phi + 4X^2} - \tan^2 \phi}{1 - \tan^2 \phi}
\ {\tanh\left( \beta H \right)} 
\label{eq:0T}
\\
&\approx&
\left \{ 
\begin{array}{ll}
\frac{1}{2} \ \frac{| \sin 2 \phi |}{|\sin \phi| + |\cos \phi |} 
\ {\tanh\left( \beta H \right)}
  & \mbox{for $\tan \phi \gg X$} \\
\\
\sqrt{\tan^2 \phi + 4X^2}
\ {\tanh\left( \beta H \right)}
  & \mbox{for $\tan \phi \ll X$} 
\end{array} 
\right. 
\label{eq:1step} 
\end{eqnarray} 
The first line of Eq.~(\ref{eq:1step}) corresponds to the single-step model 
\cite{DEVI92,SPOH93,MEAK86,PLIS87}. 
In Ref.~\cite{SPOH93} the same result was obtained from 
a Green-Kubo-like linear-response formula for the interface mobility. 
In the second line we retain only those terms in $\tan^2 \phi$, which 
dominate in the limit $|\tan \phi| \ll X$, $X \rightarrow 0$. 
It corresponds to the polynuclear growth (PNG) model 
\cite{DEVI92,KRUG89,KERT89}. Equation~(\ref{eq:0T}) provides a correct
interpolation between the PNG and single-step results as $|\tan \phi|$ 
increases from $|\tan \phi| \ll X$ to $|\tan \phi| \gg X$. 
Only the factor $\tanh (\beta H)$, which corresponds to the average 
velocity of a single step along the surface, depends 
explicitly on the specific dynamic.
At higher temperatures than those for which Eq.~(\ref{eq:0T}) holds, 
the other spin classes contribute to the interface velocity as well, and our 
approximation goes beyond the single-step and PNG approximations.

\section{Nonlinear-Response SOS Approximation} 
\label{sec:NLR} 

The velocity estimates obtained in Sec.~\ref{sec:SOS} 
were derived from the equilibrium interface fluctuations 
at $H=0$ and thus constitute a linear-response approximation. 
This is satisfactory for sufficiently weak fields. For stronger 
fields, however, the structure of the moving interface is, in general,
modified, leading to 
additional field dependences in the velocity. As will be shown in 
Sec.~\ref{sec:MC}, this effect can be significant. 
In this section we therefore develop a 
mean-field theory for the step-height pdf for a moving, flat 
SOS interface in a nonzero field. 

The step-height pdf for a {\it stationary\/} SOS interface in a 
{\it nonzero\/} field is \cite{BURT51} 
\begin{equation}
p_x[\delta(x)] = Z_x^{-1} X^{|\delta(x)|}
\ e^{ \left[ \gamma(\phi) - 2 \beta H x \right] \delta(x) } \;. 
\label{eq:step_pdfH}
\end{equation} 
The term containing $H$ adds an $x$-dependence to the Lagrange 
multiplier which determines  $\langle \delta(x) \rangle$.The corresponding 
$x$-dependence in the partition function is obtained by replacing 
$\gamma(\phi)$ by $\left[ \gamma(\phi) - 2 \beta H x \right]$ 
in Eq.~(\ref{eq:Z}), where now $\tan \phi = \langle \delta(0) \rangle$.
The geometric structure described by Eq.~(\ref{eq:step_pdfH}) 
is a macroscopically curved interface which bulges 
in the direction of the metastable phase region.
{}For the case of conserved order parameter or nonconserved order parameter 
with an interface pinned at two points, 
the stationary configuration is an equilibrium one. 
{}For nonconserved order parameter without pinning, 
it corresponds to a critical droplet of the stable phase.
If the average curvature is changed from the $x$-dependent form given by 
Eq.~(\ref{eq:step_pdfH}), the interface will move.

It is well known that the macroscopic, stationary distribution for flat,
moving interfaces in the KPZ universality class is Gaussian 
\cite{BARA95,KARD86}, corresponding to a random walk with independent
increments. Nevertheless, the step heights in several discrete models in
this class are known to be 
correlated at {\it short\/} distances \cite{NEER97,KORN00}. 
Here we develop a mean-field theory for the single-step-height
pdf of the moving interface
by ignoring these short-range correlations. Thus we assume that the step
heights are statistically independent and
identically distributed, just as they are for $H=0$. 
In this approximation, the single-step pdf of a moving
interface parallel to the $x$-axis is given by Eqs.~(\ref{eq:step_pdf}) 
and~(\ref{eq:Z}) with $\gamma =0$ and an 
$H$-dependent generalization of the parameter $X$. 
We now construct a self-consistency equation to determine this parameter,
$X(T,H)$.

\begin{figure}[ht]
\begin{center}
\includegraphics[angle=270,width=5in]{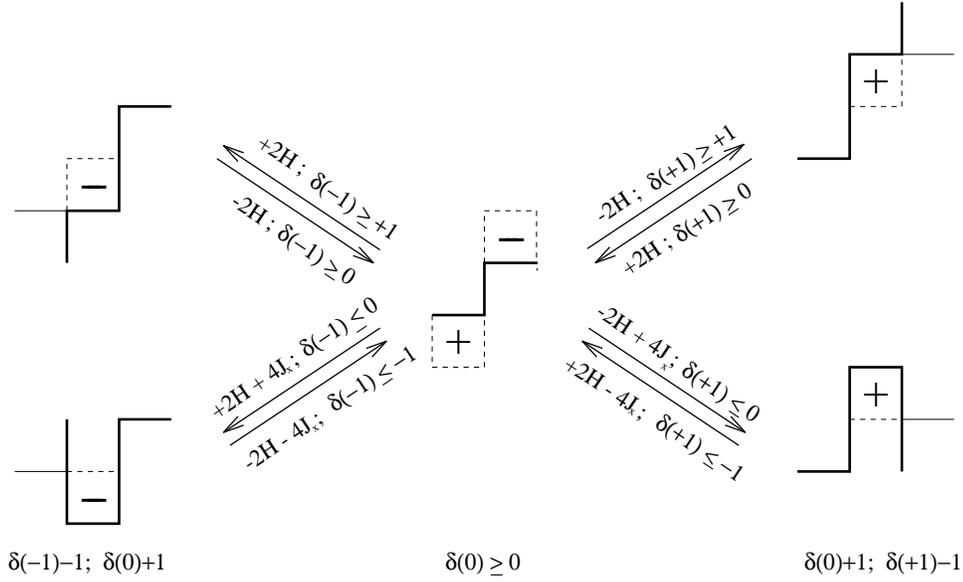}
\end{center}
\caption[]{
Figure for calculating the transition rates 
${\cal W}[\delta(0) \rightarrow \delta(0)+1]$ and 
${\cal W}[\delta(0) +1 \rightarrow \delta(0)]$, Eq.~(\ref{eq:rates}).   
Interface configurations are indicated by bold line segments. Like in 
Fig.~\protect\ref{fig:SOS}, the negatively magnetized phase is above the
interface, and the positively magnetized phase below it. 
At the center is shown a
step $\delta(0) \ge +1$ (here shown as $\delta(0) = +1$ for
concreteness). A transition to $(\delta(0)+1)$ can be effected
by flipping either one of the spins in the dashed boxes, each with
probability 1/2. 
Flipping the initially negative spin ($-$) decreases
$\delta(+1)$ by one. The resulting configurations are shown to the
right. The energy change depends on the initial
value of $\delta(+1)$, and the two possible energy changes and the
corresponding conditions are shown next to the right-pointing arrows. 
The thin horizontal lines represent the interface
configuration corresponding to the equality in the conditions. 
The energy changes and conditions for the reverse transitions are
shown next to the left-pointing arrows. 
Flipping the initially positive spin ($+$)  
analogously decreases $\delta(-1)$ by one. 
The energy changes and conditions for the forward and reverse
transitions resulting from flipping this spin, 
are indicated in the left-hand half of the figure. 

}
\label{fig:rate}
\end{figure}

{}For simplicity we consider the step at $x=0$, and we
take $\delta(0) \ge 0$.  The equations obtained by considering 
other values of $x$ and 
$\delta(x) \le 0$ are identical to the one we derive below. 
The total transition probability for the height of a single step 
to change from $\delta$ to $\delta\pm 1$
is ${\cal W}[\delta \rightarrow \delta \pm1]$. Relating the single-step
transition probability to the single-step-height pdf by  
detailed balance, we have:  
\begin{equation}
X(T,H) 
\equiv 
\frac{p_0[\delta(0)+1]}{p_0[\delta(0)]} 
=
\frac{{\cal W}[\delta(0) \rightarrow \delta(0)+1]}
{{\cal W}[\delta(0)+1 \rightarrow \delta(0)]}
\label{eq:step_1}\;. 
\end{equation} 
To find these transition probabilities, 
we refer to Fig.~\ref{fig:rate}. In order for  
$\delta(0)$ to increase by one, either  
the spin at $x = +1/2$ can flip from $-1$ to $+1$, decreasing 
$\delta(+1)$ by one, or 
the spin at $x = -1/2$ can flip from $+1$ to $-1$, decreasing 
$\delta(-1)$ by one. Each of these possibilities is attempted with 
probability 1/2. For each spin-flip direction, the resulting energy
change can have two different values, depending
on the height of the corresponding neighboring step. 
Analogous arguments hold for the reverse transitions, $\delta(0)+1 \rightarrow
\delta(0)$. 
The energy changes and the conditions on the neighboring step heights
are given next to the arrows denoting the directions of the transitions
in Fig.~\ref{fig:rate}. 
Expressing the single-spin transition rate corresponding to an energy
change $\Delta E$ as $W[\Delta E]$ and reading the energy changes and
conditions off from Fig.~\ref{fig:rate}, we get:
\begin{eqnarray}
{\cal W}[\delta(0) \rightarrow \delta(0)+1] 
&=& 
\frac{1}{2} \Big\{ 
W[-2H]P[\delta(+1)\ge+1] + 
W[-2H+4J_x]P[\delta(+1)\le0] 
\nonumber\\
& & + 
W[+2H]P[\delta(-1)\ge+1] + 
W[+2H+4J_x]P[\delta(-1)\le0]  
\Big\}
\nonumber\\
{\cal W}[\delta(0)+1 \rightarrow \delta(0)] 
&=& 
\frac{1}{2} \Big\{ 
W[+2H]P[\delta(+1)\ge0] + 
W[+2H-4J_x]P[\delta(+1)\le-1]  
\nonumber\\
& & + 
W[-2H]P[\delta(-1)\ge0] + 
W[-2H-4J_x]P[\delta(-1)\le-1]  
\Big\}
\nonumber\\
\label{eq:rates}
\end{eqnarray}
Consistent with the mean-field approximation we 
calculate the cumulative probabilities with the same stationary 
single-step pdf, obtaining 
$P[\delta \ge 0] = P[\delta \le 0] = 1/[1+X(T,H)]$ 
and
$P[\delta \ge +1] = P[\delta \le -1] = X(T,H)/[1+X(T,H)]$. 
The resulting self-consistency equation for $X(T,H)$ is 
\begin{equation}
\label{eq:sceq}
X(T,H) = 
\frac{X(T,H) \left\{ W[-2H]+W[+2H]\right\} + W[-2H+4J_x] + W[+2H-4J_x]}
{
W[-2H]+W[+2H] + X(T,H) \left\{ W[-2H-4J_x]+W[+2H-4J_x]\right\}
} 
\;.
\end{equation}
This is solved to yield
\begin{equation}
X(T,H) = e^{-2 \beta J_x} 
\left\{
\frac{e^{-2 \beta H}W[-2H-4J_x] + e^{2\beta H}W[+2H-4J_x]}
{W[-2H-4J_x] + W[+2H-4J_x]}
\right\}^{1/2}
\;,
\label{eq:XTH}
\end{equation}
where we have also used the detailed-balance relation for the
single-spin transition probabilities, 
$W[+\Delta E]/W[-\Delta E] = e^{- \beta \Delta E}$, 
to eliminate $W[+2H+4J_x]$ and $W[-2H+4J_x]$. 

The approach described above is also applicable to curved interfaces, 
and the resulting self-consistency equations are cubic in $X(T,H)$. 
However, except for the stationary curved interface described by 
Eq.~(\ref{eq:step_pdfH}), the shapes of such interfaces are not stationary 
and much more difficult to investigate by simulations. 
We hope to return to these more complicated problems in the future. 

Equation~(\ref{eq:XTH}) shows that $X(T,H)$ depends on the specific 
dynamic, except for $H=0$, where it reduces to its equilibrium value, 
$X(T,0) = e^{-2 \beta J_x}$, for all dynamics that satisfy detailed 
balance.

\begin{figure}[ht]
\begin{center}
\includegraphics[angle=270,width=5in]{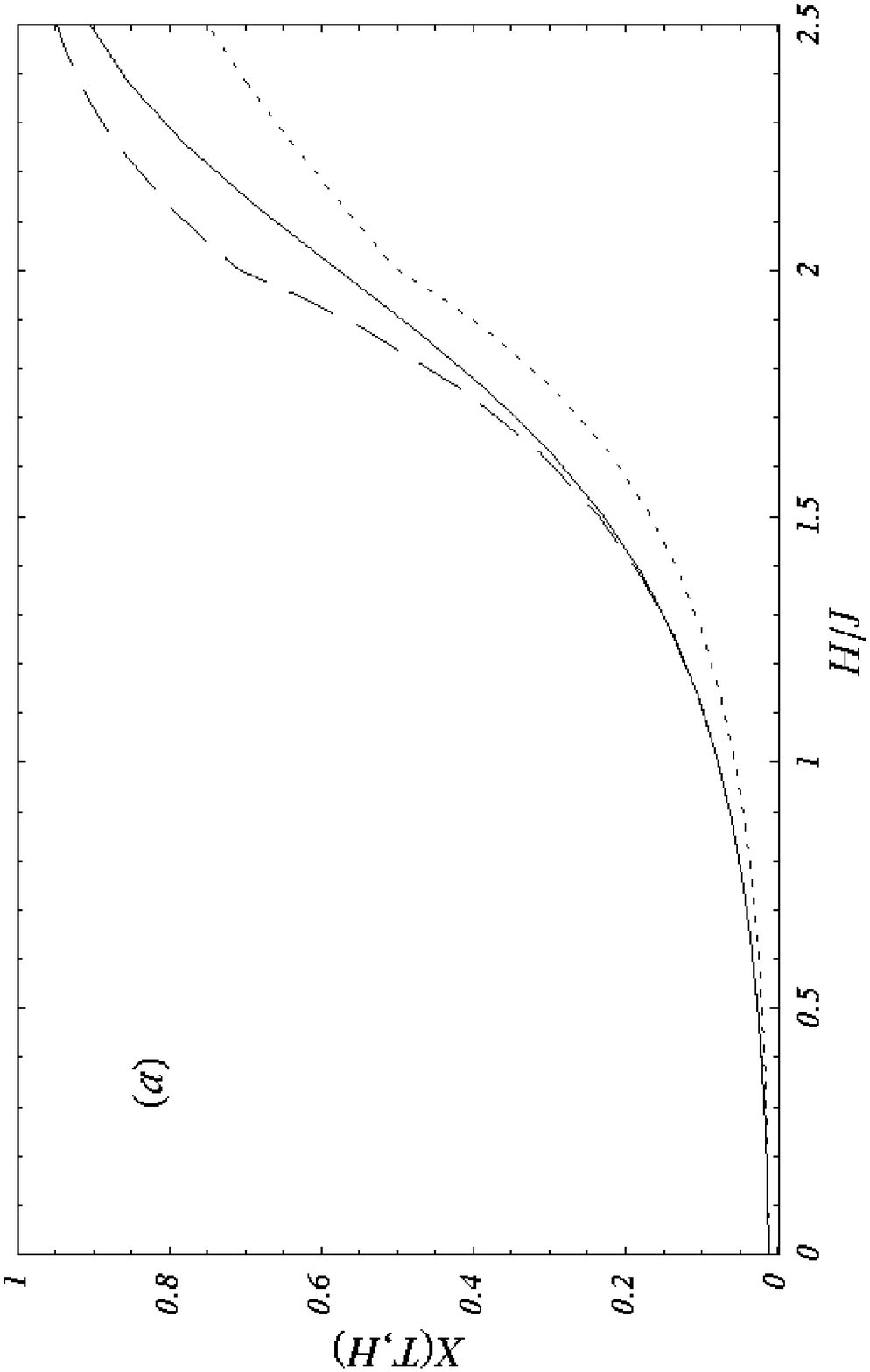}
\end{center}
\begin{center}
\includegraphics[angle=270,width=5in]{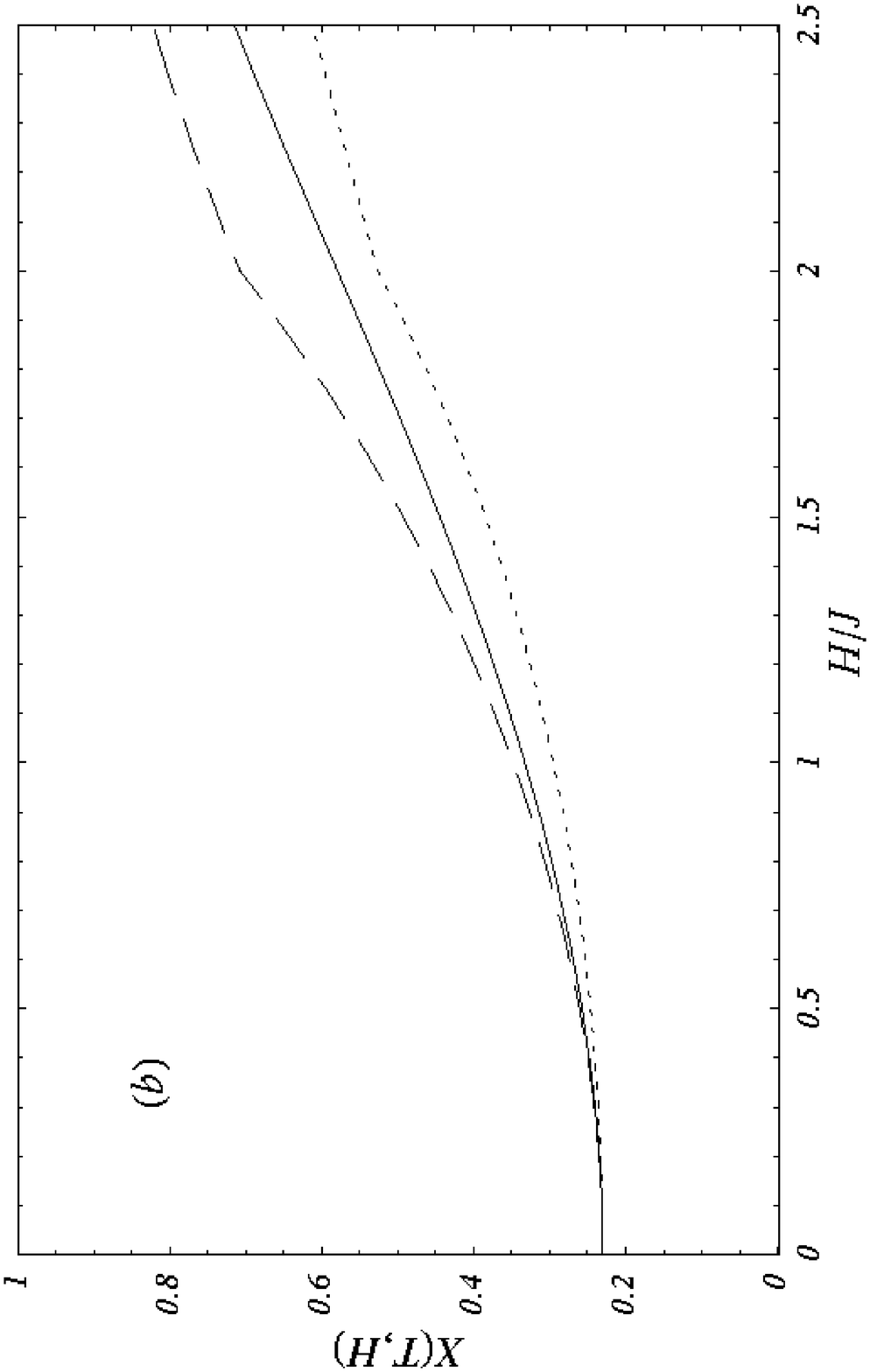}
\end{center}
\caption[]{
The parameter $X(T,H)$ for different dynamics, shown vs $H$ at 
$T=0.2T_c$ (a) and $T=0.6T_c$ (b). 
Solid curves: Glauber dynamic, $X_{\rm G}(T,H)$, 
Eq.~(\protect\ref{eq:XG}). 
Long-dashed curves: Metropolis dynamic, $X_{\rm M}(T,H)$,
Eq.~(\protect\ref{eq:XM}). 
Dotted curves: The approximation used in Ref.~\protect\cite{RAMO99}. 
}
\label{fig:XGM}
\end{figure}

Using the Glauber dynamic defined by Eq.~(\ref{eq:glau}), we get 
explicitly 
\begin{equation}
X_{\rm G}(T,H) = 
e^{-2 \beta J_x} 
\left\{
\frac{e^{2 \beta J_x} \cosh(2 \beta H) + e^{-2\beta J_x}}
{e^{-2 \beta J_x} \cosh(2 \beta H) + e^{2\beta J_x}}
\right\}^{1/2}
\;.
\label{eq:XG}
\end{equation}
The Metropolis dynamic, 
$W_{\rm M}[\Delta E] = \min \left[1,e^{- \beta \Delta E}\right]$, yields 
\begin{equation}
X_{\rm M}(T,H)  
= 
\left \{ 
\begin{array}{ll}
e^{-2 \beta J_x} \left\{ \cosh (2 \beta H) \right\}^{1/2}
  & \mbox{for $H \le 2J_x$} \\
\\
\left\{
\frac{1+ e^{- \beta (2H+4J_x)}}
{e^{- \beta (2H-4J_x)}+1}
\right\}^{1/2}
  & \mbox{for $H \ge 2J_x$} \\
\end{array} 
\right. 
\;.
\label{eq:XM}
\end{equation}
Numerically, $X_{\rm G}$ and $X_{\rm M}$ are not very different, 
and they both approach unity from below 
as $H$ increases beyond $2J_x$. They are shown together vs $H$ in 
Fig.~\ref{fig:XGM}. 

The approximation used in Ref.~\cite{RAMO99}, 
$X(T,H) = e^{-2 \beta J_x} \cosh(\beta H)$, can be obtained by 
``brutally decoupling'' 
the steps through fictitiously splitting each spin in two and 
flipping only half of a spin to change the height of a single step. 
{}For such a process Eq.~(\ref{eq:step_1}) directly yields 
$X(T,H) = \{W[-H+2J_x]+W[+H+2J_x]\}/\{W[+H-2J_x]+W[-H-2J_x]\}$, from which 
the result in Ref.~\cite{RAMO99} is obtained for $H < 2J_x$ using the 
Metropolis dynamic. It is surprisingly close to the proper mean-field results  
and is included in Fig.~\ref{fig:XGM} for comparison. 

In their study of a lattice-gas model for three-dimensional crystal growth 
\cite{KOTR91}, Kotrla and Levi introduced a single-site dynamic which in Ising 
language can be described (up to an overall rate factor $e^{2 \beta H}$) 
as the product of Metropolis for field effects and 
Glauber for interaction effects: 
$W_{\rm KL}[-2H-4J_x] = W_{\rm G}[-4J_x]$ and 
$W_{\rm KL}[+2H-4J_x] = e^{-2 \beta H} W_{\rm G}[-4J_x]$. 
Inserting these transition probabilities in Eq.~(\ref{eq:XTH}), one 
finds that they lead to an $H$-independent $X$. This is indeed the case 
for any dynamic in which the field and interaction effects are 
statistically independent and obey detailed balance separately. It shows that 
the local interface structure of driven interfaces can vary strongly 
as the dynamics are changed, even within the class of 
nonconservative single-spin-flip dynamics. 
Dynamics that factorize in this way
are known as ``soft,'' in contrast to the ``hard''
dynamics which do not factorize, such as Metropolis 
and the Glauber rate used here. 
Soft and hard dynamics have been shown to lead to
differences in the steady states in a number of other nonequilibrium
systems as well \cite{MARR99}. 

All the results for the spin-class populations of the 
zero-field equilibrium interface, which were derived in Sec.~\ref{sec:SOS} 
and are listed in Table~2, 
can now be applied to obtain a nonlinear-response approximation for the 
propagation velocity of flat, driven interfaces. 
This simply requires replacing the zero-field $X=e^{-2 \beta J_x}$ used in the 
linear-response approximation by the 
field-dependent $X(T,H)$, obtained from Eq.~(\ref{eq:XTH}) with the 
transition probabilities corresponding to the particular dynamic used. 
{}For most hard dynamics, the net effect is to increase the mean step 
height, $\langle | \delta | \rangle$, and thus the interface velocity. 

In the next section we show that this nonlinear-response approximation 
gives good agreement with MC simulations of driven, flat Ising interfaces 
for a wide range of fields and temperatures.

\section{Comparison with Monte Carlo Simulations} 
\label{sec:MC} 

We have compared the analytical estimates of propagation velocities 
and spin-class populations developed above with MC simulations of the 
same model for $J_x = J_y = J$. 

\subsection{Simulation details}
\label{sec:MCs}
To minimize the finite-size effects (see below), large 
simulation lattices are needed to accommodate
a sufficient length of the interface and provide enough room for 
it to move unimpeded. To achieve long 
simulation runs it is necessary to employ a co-moving
simulation box to prevent the interface  from hitting the 
system boundary in the $y$-direction. We therefore used 
an ``active-zone''  algorithm which relies on the fact that spins 
with no broken bonds have zero transition probability.  
The algorithm uses a lattice of size $L_x$ in the $x$-direction, 
and it keeps a list of all spins which have at least
one broken bond and thus can flip in the next simulation step.
Once a spin loses its last broken bond, 
it is removed from the ``active list.''  
A new spin is added to the list as soon as it acquires a broken bond 
due to a transition of one of its neighbors. 
The memory required for the list is proportional to the length of the
interface, $L_x / |\cos \phi |$. 
No lattice boundaries are needed in the $y$-direction, and consequently 
arbitrarily long simulation runs can be performed. 
Except for these modifications, the algorithm is a 
straightforward implementation of the discrete-time 
$n$-fold way \cite{BORT75}. 

The mean tilt angle $\phi$ was fixed by helical boundary 
conditions in the $x$-direction. The production runs were performed with  
$L_x = 1000$ and fixed $\phi$ between 0 and $\pi/4$. Since kinetic Ising 
interfaces belong to the KPZ universality class 
\cite{DEVI92,KARD86}, the macroscopic interface width, 
$\langle \left[ y(x,t) - \langle y(x,t) \rangle \right]^2 \rangle^{1/2}$,
is expected to saturate at a value proportional to $L_x^{1/2}$ after a 
time $t \propto L_x^{3/2}$. 
In order to ensure stationarity we ran the simulation for 10\,000~MCSS 
before taking any measurements.  Exploratory runs with both 
larger and smaller $L_x$ and 
``warm-up'' times showed that the values used in the 
production runs were 
sufficient to ensure a macroscopically stationary interface. 
Class populations and 
interface velocities were averaged over 200\,000~MCSS. 
The macroscopic stationarity of the interface should abundantly 
ensure the stationarity of these 
quantities, which are properties of the {\it local\/} interface structure. 

\subsection{Interface velocities}
\label{sec:MCi}
\begin{figure}[ht]
\begin{center}
\includegraphics[angle=270,width=5in]{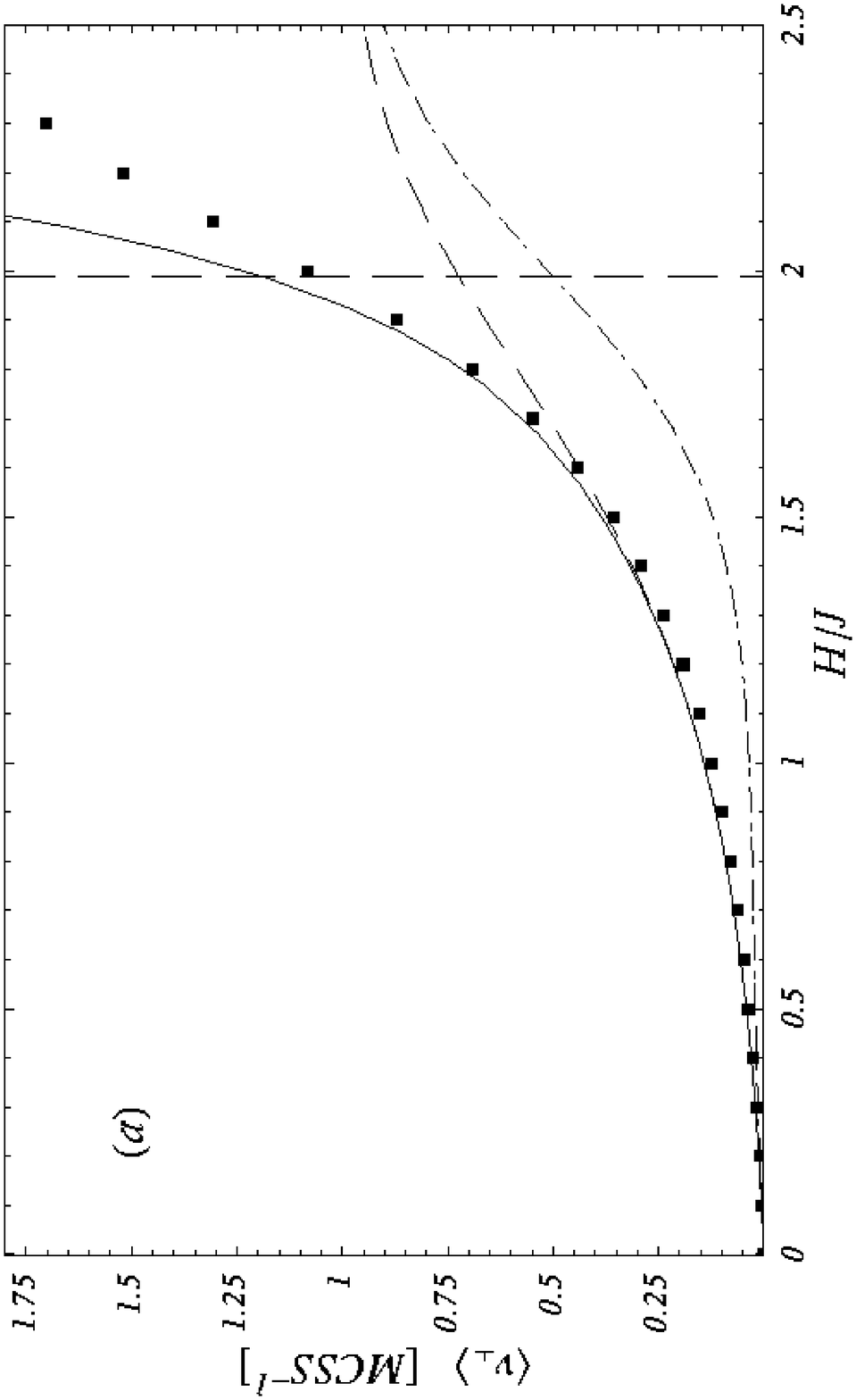}
\end{center}
\begin{center}
\includegraphics[angle=270,width=5in]{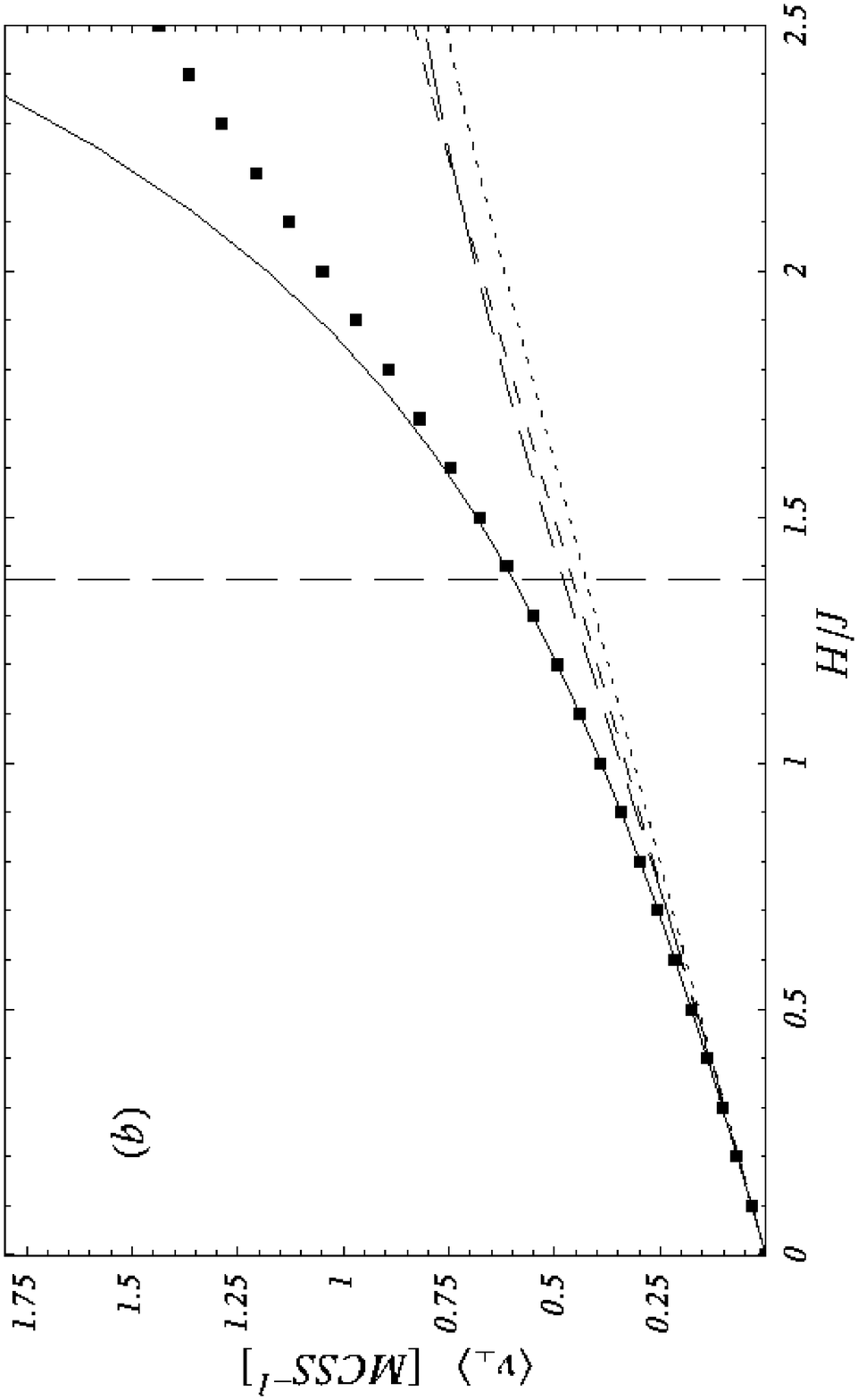}
\end{center}
\caption[]{
Comparison between MC simulations and the four different 
analytic approximations introduced 
here. Shown is the normal interface velocity $\langle v_\perp \rangle$ vs
$H$ for $\phi = 0$ at $T = 0.2T_c$ (a) and $T=0.6T_c$ (b). 
MC data (squares), 
linear-response excluding SOS-violating transitions (dotted), 
linear-response including SOS-violating transitions (short-dashed), 
nonlinear-response excluding SOS-violating transitions (long-dashed), 
and nonlinear-response including SOS-violating transitions (NLI) (solid). 
In the low-temperature case shown in (a)
the two linear-response approximations are almost indistinguishable, 
and the dotted and short-dashed curves coincide as the lowest of the 
three curves shown. 
Statistical errors in the MC data are everywhere
much smaller than the symbol size, both in this and all subsequent figures. 
The vertical long-dashed lines mark the crossover field 
$H_{\rm MFSP}(T)$. 
In both (a) and (b), the NLI approximation gives the best overall 
agreement with the data. 
It is the only one which will be used in subsequent figures. 
}
\label{fig:veloH}
\end{figure}

The overall quality of the four different approximations developed above 
(linear-response (Sec.~\ref{sec:SOS}) 
and nonlinear-response (Sec.~\ref{sec:NLR}), 
each either excluding or including the SOS-violating 
transitions from classes $10s$ and $20s$) were explored in a wide 
range of $H$ and $T$.  
In Fig.~\ref{fig:veloH} the four approximations are 
compared with MC data for interfaces parallel to the $x$-axis at 
$T = 0.2T_c$ [Fig.~\ref{fig:veloH}(a)] 
and $T = 0.6T_c$ [Fig.~\ref{fig:veloH}(b)], 
where $T_c$ is the exact critical temperature. 
The crossover field $H_{\rm MFSP}(T)$ is marked by a vertical dashed 
line in both (a) and (b). 
At the lowest temperature, the interface has very few steps
higher than one for $H=0$. 
As a result, it makes practically no difference whether
SOS-violating transitions are allowed or not, and the curves
representing the linear-response approximations including and excluding 
SOS-violating transitions coincide in Fig.~\ref{fig:veloH}(a). 
The nonzero field increases the average step height,
and as a result the SOS-violating transitions contribute significantly
for stronger fields in the nonlinear-response approximation. 
Overall, the nonlinear-response approximation including SOS-violating 
transitions (``nonlinear inclusive'' or NLI) 
is everywhere better than the others and is particularly 
good for $H < H_{\rm MFSP}(T)$. 
It is the only one which will be used in the rest of this paper. 

\begin{figure}[ht]
\begin{center}
\includegraphics[angle=270,width=5in]{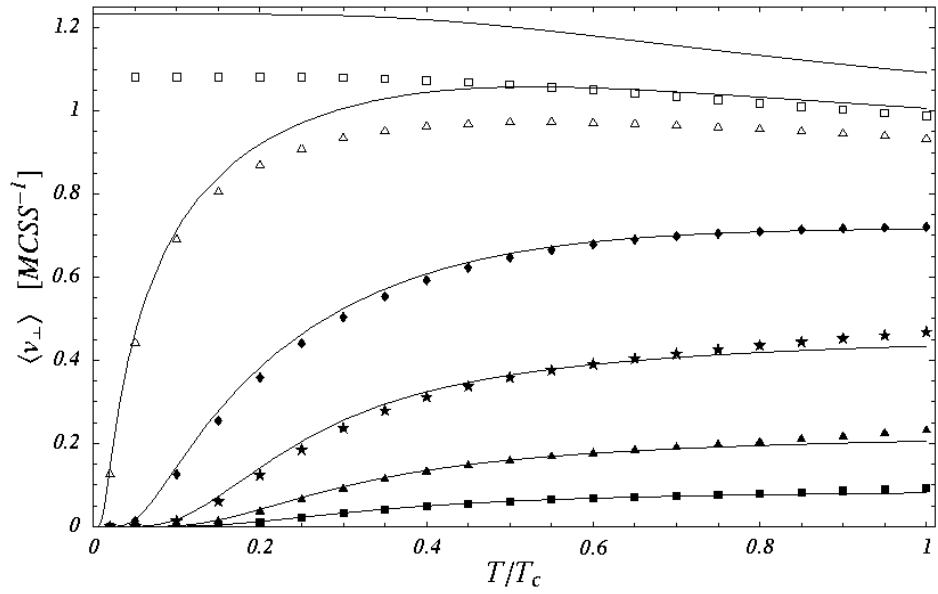}
\end{center}
\caption[]{
Temperature dependence of the normal interface velocity, 
$\langle v_\perp \rangle$, for $\phi=0$ 
at $H/J = 2.0$, 1.9, 1.5, 1.0, 0.5, and 0.2 from top to bottom. 
MC simulations (data points) 
and the NLI analytic approximation (solid curves). 
}
\label{fig:veloT}
\end{figure}

The temperature dependence between $T=0$ and $T_c$ 
of the velocities of interfaces parallel to the $x$-axis are shown in 
Fig.~\ref{fig:veloT} for several values of $H/J$ between 0.2 and 2.0. 
{}For $H/J \le 1.5$, the discrepancy is nowhere greater than 
a few percent -- mostly much smaller.  Even for $H/J=1.9$ and 2.0, the
discrepancy remains below about 15\% everywhere. 

\begin{figure}[ht]
\begin{center}
\includegraphics[angle=270,width=5in]{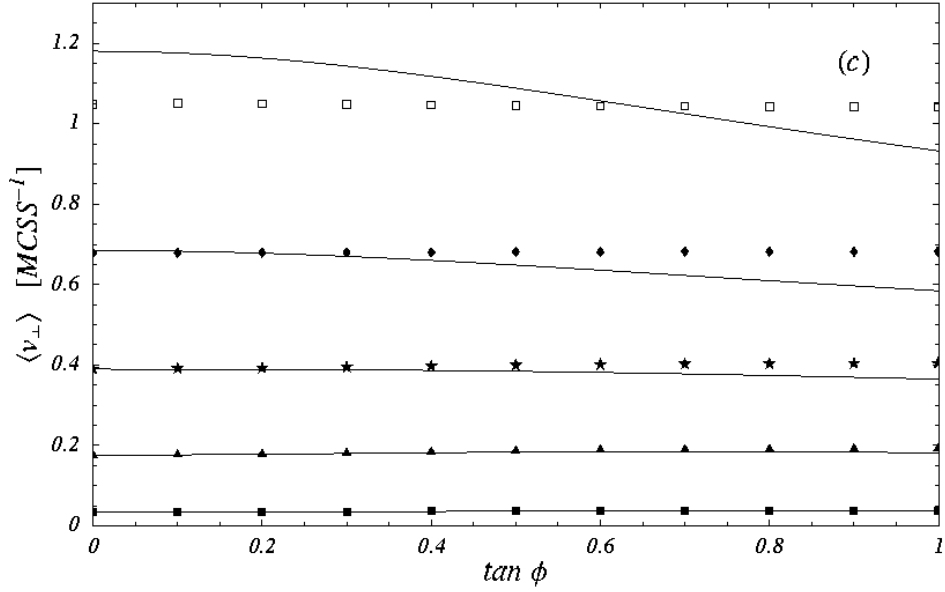}
\end{center}
\caption[]{
The normal interface velocity $\langle v_\perp \rangle$, shown  
vs $\tan \phi$ for different fields.
Simulation data are shown as data points and the NLI approximation as 
solid curves. The fields included are 
(from top to bottom) $H/J = 2.0$ (empty squares, only in a and c), 
1.9 (empty triangles, only in a), 1.5 (filled diamonds), 
1.0 (filled stars), 0.5 (filled triangles), and 0.1 (filled squares). 
(a): At $T=0.2T_c$. 
(b): The four weakest fields 
at $T=0.2T_c$, divided by $\tanh(\beta H)$. This shows the crossover
between PNG and single-step growth, Eqs.~(\protect\ref{eq:0T})
and~(\protect\ref{eq:1step}). The dotted curve is the
zero-temperature single-step result.
(c): At $T=0.6T_c$. 
}
\label{fig:veloA}
\end{figure}

The anisotropy of the interface velocities is shown in Fig.~\ref{fig:veloA} 
for several values of $H/J$ between 0.1 and 2.0. 
The agreement between the  NLI approximation and the simulations is very 
good, except for the strongest fields. 
At $T=0.2T_c$ 
[Fig.~\ref{fig:veloA}(a)], 
both the analytic and simulation results for $H/J \le 1.5$ increase with 
increasing $\phi$. As predicted by Eqs.~(\ref{eq:0T}) and~(\ref{eq:1step}), 
under weak fields they cross over from the form of the PNG model 
for $\tan \phi \ll X$, to almost exact agreement with the 
single-step model for $\tan \phi \gg X$ [Fig.~\ref{fig:veloA}(b)]. 
Growth shapes generated from these velocities 
would be almost square, with their sides parallel to the $x$ and $y$ axes 
\cite{DEVI92}. For strong fields, however, the analytic approximation 
predicts velocities that are slightly larger 
in the symmetry directions than for inclined interfaces, as is the case 
for the Eden model \cite{DHAR86,MEAK86B,HIRS86}. 
This reverse anisotropy is not seen in the MC data, which become 
almost perfectly isotropic and would lead to circular growth shapes. 
With the Glauber dynamic it appears that stronger fields and lower 
temperatures are 
needed to observe reverse anisotropy in the MC data \cite{DEVI92}. 
However, the discrepancies between the simulations and 
the analytic approximation are modest in the regime explored here, 
even near $\phi = \pi/4$ for strong fields. For small $\phi$ the 
agreement remains good for all fields $H/J \le 1.9$. 
At $T=0.6T_c$ [Fig.~\ref{fig:veloA}(c)], 
the simulated velocities are practically isotropic for all fields. 
The analytic NLI approximation works well, except for gradually increasing  
Eden-like reverse anisotropy for the stronger fields. 

\subsection{Spin-class populations}
\label{sec:MCc}
\begin{figure}[ht]
\begin{center}
\includegraphics[angle=270,width=5in]{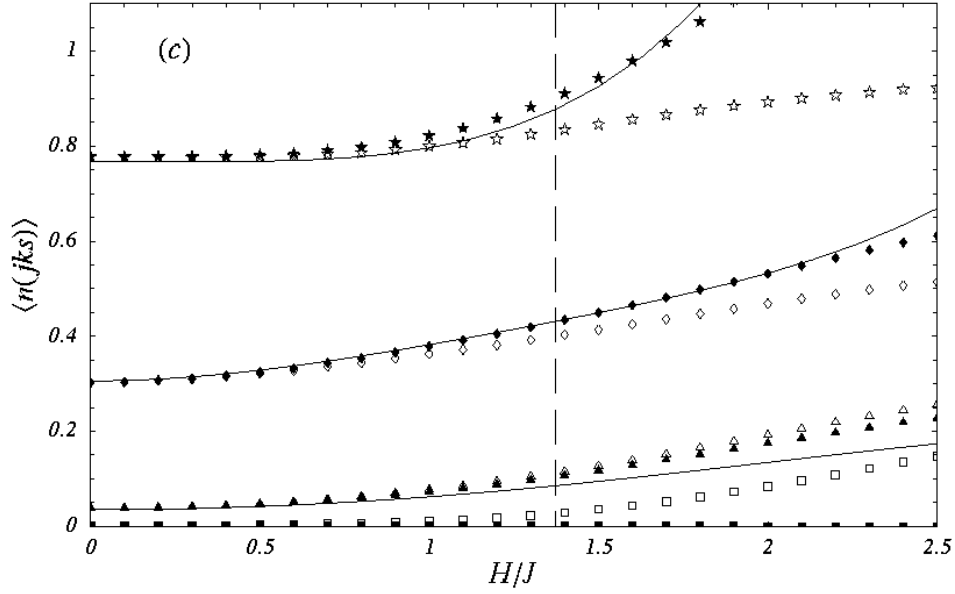}
\end{center}
\caption[]{
Comparison between MC simulations 
(filled symbols for $s=+1$ and empty symbols for $s=-1$) 
and the NLI analytic approximation (solid curves) for the spin-class 
populations. Results for $T=0.6T_c$ and $\phi=0$ are shown vs $H$. 
(a): 
Individual populations of the spin classes compatible with the SOS picture. 
From top to bottom in the left part of the 
figure are $01s$ (stars), $11s$ (diamonds), $10s$ (stars), 
$21s$ (triangles), and $20s$ (diamonds). 
(b): 
Individual populations of the spin classes that would be unpopulated for a 
perfect SOS surface: 
$02s$ (diamonds), $12s$ (triangles), and $22s$ (squares). 
These nonzero populations clearly indicate the gradual breakdown of the SOS 
approximation with increasing $H$. 
(c): 
Class populations combined according to number of broken bonds, $j+k$. 
From top to bottom: one (stars), two (diamonds), three (triangles), and 
four (squares) broken bonds. 
}
\label{fig:class}
\end{figure}

While our main emphasis is on estimates of the interface velocity, the 
most detailed information about the strengths and weaknesses of our 
mean-field approximation for the interface structure
is found in the individual spin-class 
populations. Examples of the $H$-dependences for 
$\phi = 0$ at $T = 0.6T_c$ are 
shown in Fig.~\ref{fig:class}. Individual populations are shown in 
Figs.~\ref{fig:class}(a) and (b), while spin classes with the 
same number of broken bonds are combined 
in Fig.~\ref{fig:class}(c). The crossover field 
$H_{\rm MFSP}(T)$ is shown in all three panels as a vertical dashed
line. The agreement between the 
theoretical estimates and the MC data is good out to 
about this field. 

The discrepancies that develop with increasing $H$ stem from two
sources. One is the fact that the simulated interface, unlike the SOS
description used in the theoretical analysis, is not restricted
to be free of overhangs and bubbles. The other is the development of
correlations between neighboring step heights, which
are not included in our single-step
mean-field theory for the local structure of the driven SOS interface. 

The presence of overhangs and bubbles is 
clearly reflected in the increase of $\langle n(01s) \rangle$, relative to 
the monotonically decreasing analytic approximation 
[Fig.~\ref{fig:class}(a)],
and in the nonzero populations of the classes that are not populated in the 
SOS approximation [Fig.~\ref{fig:class}(b)]. 

The increasing correlations between nearest-neighbor steps is expressed
by the gradual disappearance of the symmetry between the 
populations of spins with $s = +1$ (stable) and $s= -1$ (unstable) 
as the field is increased. The
effect of these correlations is typically to broaden protrusions on
the leading side of the interface (``hilltops'') and sharpen
protrusions on the trailing side (``valley bottoms'')
\cite{NEER97}, or {\it vice versa\/}  
\cite{KORN00}. 
In terms of spin-class populations, the former corresponds to
$\langle n(21+) \rangle < \langle n(21-) \rangle$
and
$\langle n(11+) \rangle > \langle n(11-) \rangle$. 
This is precisely what is seen in the simulations
[Fig.~\ref{fig:class}(a)]. 
More generally, we find that the 
populations of $s=+1$ are enhanced for the classes 
with one and two broken bonds, while the populations of $s=-1$ are enhanced 
for the classes with three and four broken bonds
[Fig.~\ref{fig:class}(c)]. The latter include both sharp valley bottoms
and bubbles of the unstable phase, which form a wake behind the moving 
interface. This shows that the breakdown of the SOS description and the
evolution of lateral correlations are not independent at strong fields.
The overall picture is essentially the same at $T=0.2T_c$:  
both the populations in the non-SOS classes, and the asymmetry become  
significant only near $H_{\rm MFSP}(T)$, which at that temperature is close 
to $H/J =2$.

\section{Discussion}
\label{sec:DISC}

In this paper we introduced and explored analytic approximations for 
the propagation velocity and spin-class populations
of a field-driven interface in a two-dimensional kinetic Ising model. 
The model is applicable to a number of systems undergoing order-disorder 
phase transformations, including magnetic and ferroelectric systems and 
crystal growth under conditions which are not limited by diffusion. 

The approximations are based on a linear-response approach, in which the 
equilibrium fluctuations of the interface (as embodied in average 
spin-class populations) were estimated from the Burton-Cabrera-Frank 
SOS model. The theory was extended by developing a mean-field
approximation for the 
field-dependent spin-class populations in a moving flat 
interface, yielding a nonlinear-response approximation for the 
interface velocity. This extension 
considerably improves the agreement with MC simulations. 

Our simulation results are consistent with those of Ref.~\cite{DEVI92}. 
However, since that study used the Metropolis dynamic [see their Eq.~(3)], 
the velocities cannot be compared directly. For asymptotically low $T$, 
the velocity is completely determined by the kink density, 
$\langle n(11s) \rangle$. This is the regime in which the single-step and PNG 
models are expected to hold. Indeed, they emerge in the proper limits 
from our analytic approximation, as shown in 
Eqs.~(\ref{eq:0T}) and~(\ref{eq:1step}). 
For larger $H \le 2J$ and $T \le T_c$, other spin classes also contribute 
significantly to the interface velocity. 
In this parameter range we obtain good agreement between theory and simulation 
almost everywhere. This agreement includes the disappearance of the 
velocity anisotropy with increasing $T$ and $H$. 

Both the SOS approximation for the interface
structure, and the assumption of uncorrelated step heights employed in
the nonlinear-response approximation break down for stronger fields. 
While the NLI approximation for the interface velocities 
remains very satisfactory overall, the breakdown of these assumptions
can be detected more clearly in the spin-class populations. The only 
detailed theories of the local structure of driven interfaces that we
are aware of, are for
much simpler SOS models than the unrestricted one studied here, such as the 
body-centered SOS (BCSOS) and restricted SOS (RSOS) models \cite{NEER97}. 
It would therefore be interesting to compare our 
mean-field approximation for the interface structure with MC
simulations of the driven SOS interface, so that the complications
arising from overhangs and bubbles can be avoided \cite{NEW}. 

In summary, the approximations presented here give results for the spin-class 
populations and propagation velocities of Ising interfaces driven by 
nonzero fields, which are exact in the asymptotic limits of low $T$ and $H$, 
and which agree well  
with MC simulations almost everywhere, even for $H$ near $2J$ 
and $T$ near $T_c$. It is not clear whether the wide regime of applicability 
is an accident limited to two dimensions, and it would therefore be 
interesting to extend the approach to three dimensions as well.

\section*{Acknowledgments}
\label{sec:ACK}

P.~A.\ Rikvold dedicates this paper to Professor George Stell 
on his 65th birthday. His friendship, encouragement, and scientific 
guidance are deeply appreciated. 

We acknowledge useful conversations with R.~K.~P.\ Zia, 
G.~Korniss, and M.~A.\ Novotny, 
and comments on the manuscript by G.~Korniss, M.~A.\ Novotny, 
G.~Brown, and two anonymous referees. Parts of this work were performed at 
the Colorado Center for Chaos and Complexity, 
University of Colorado at Boulder; 
the Faculty of Integrated Human Studies, 
Kyoto University; 
and the Department of Physics, 
University of Texas at Austin. 
Hospitality and partial support extended to P.~A.~R.\ 
at these institutions are gratefully acknowledged. 
The research was supported in part 
by National Science Foundation Grants No.~DMR-9634873 and DMR-9981815,
and by Florida State 
University through the Center for Materials Research and Technology and 
the Supercomputer Computations Research Institute 
(U.S.\ Department of Energy Contract No.~DE-FC05-85ER25000).


\clearpage


\begin{table}
 \caption[]{
The spin classes in the anisotropic square-lattice Ising model. 
The first column gives the class labels, $jks$. 
The second column gives the total 
field and interaction energy for a spin in each class, $E(jks)$, relative 
to the energy of the state with all spins parallel and $H=0$, 
$E_0 = -2(J_x + J_y)$. 
The third column gives the change in the total system energy that would 
result from reversing a spin in a particular class from $s$ to $-s$, 
$\Delta E(jks)$. 
In both $E(jks)-E_0$ and $\Delta E(jks)$, the upper sign corresponds to $s=-1$,
and the lower sign corresponds to $s=+1$. 
The first three classes (marked  $\ast$) have nonzero populations in the 
SOS approximation, and flipping a spin in any of them 
preserves the SOS interface configuration. 
The next two classes (marked $\dag$) also have nonzero populations in the 
SOS approximation, but flipping a spin in any of them may produce an overhang 
or a bubble. The two classes marked $\ddag$ have zero populations in the 
SOS approximation, but flipping a spin in any of them may lead to a 
configuration compatible with the SOS constraint. 
Class $22s$ represents a single spin which is antiparallel to all its 
neighbors; flipping such a spin yields a bulk spin in class 00$-s$. 
Although only the classes marked $\ast$ and $\dag$ have nonzero populations in 
the SOS approximation, the MC transition probabilities of all classes 
except 00$s$ are given by Eq.~(\ref{eq:glau}). 
The bulk spins, 00$s$, have zero transition probabilities 
in the dynamic used here. 
}
\begin{center}
\normalsize
\begin{tabular}{| l | l | l | }
 \hline\hline
Class, $jks$
& $E(jks) - E_0$
& $\Delta E(jks)$
\\
 \hline\hline
 $01s$ $\ast$
 & $\pm H + 2J_y$
 & $\mp 2H + 4J_x $
\\
 \hline
 $11s$ $\ast$
 & $\pm H + 2(J_x+J_y)$
 & $\mp 2H $
\\
 \hline
 $21s$ $\ast$
 & $\pm H + 2(2J_x+J_y)$
 & $\mp 2H - 4J_x $
\\
 \hline\hline
$10s$  $\dag$
 & $\pm H + 2J_x$
 & $\mp 2H + 4J_y $
\\
 \hline
$20s$  $\dag$
 & $\pm H + 4J_x$
 & $\mp 2H - 4(J_x-J_y) $
\\
 \hline\hline
$12s$  $\ddag$
 & $\pm H + 2(J_x + 2 J_y)$
 & $\mp 2H - 4J_y $
\\
 \hline
$02s$  $\ddag$
 & $\pm H + 4J_y$
 & $\mp 2H + 4(J_x-J_y) $
\\
 \hline\hline
$22s$
 & $\pm H + 4(J_x + J_y)$
 & $\mp 2H - 4(J_x + J_y) $
\\
 \hline\hline
$00s$
 & $\pm H$
 & $\mp 2H + 4(J_x + J_y) $
\\
 \hline\hline
 \end{tabular}
 \end{center}
\label{table:class}
\end{table}

\clearpage

\begin{table}
 \caption[]{
The mean populations for the spin classes included in the SOS 
approximation, together with their contributions to the interface velocity 
under the Glauber dynamic. The first column gives the class labels, $jks$. 
The second column gives the mean spin-class populations for general tilt angle 
$\phi$, with $\cosh \gamma(\phi)$ from Eq.~(\ref{eq:chgam}). 
The third column gives the spin-class populations for $\phi = 0$. 
Using $X = e^{-2 \beta J_x}$ in these expressions, one obtains the 
linear-response result in which the spin-class populations are 
evaluated for $H=0$. Using $X = X(T,H)$ from
Eq.~(\ref{eq:XTH}) with the transition probabilities of the
particular dynamic used [here: Glauber, Eq.~(\ref{eq:XG})], 
one gets the nonlinear-response approximation,  
which includes an estimate of the field-dependent modifications 
of the spin-class populations in the moving interface. 
The fourth column contains the contributions to the mean interface 
velocity in the $y$-direction from spins in classes $jk-$ and $jk+$, 
using the Glauber dynamic. These are the only quantities in this table 
that depend explicitly on the 
specific dynamic, and they could easily be replaced with results for, e.g., 
Metropolis. 
}
\begin{center}
\normalsize
\begin{tabular}{| l | l | l | l |}
 \hline\hline
Class, $jks$
& $\langle n(jks) \rangle$, general $\phi$
& $\langle n(jks) \rangle$, $\phi=0$
& $\langle v_y(jk) \rangle$
\\
 \hline\hline
 $01s$ $\ast$
 & $\frac{1 - 2X \cosh \gamma (\phi) + X^2}{(1-X^2)^2}$
 & $\frac{1}{(1+X)^2}$
 & $\frac{\tanh\left( \beta H \right)}
         {1 + \left[\frac{\sinh \left(2 \beta J_x \right) }
                   {\cosh \left( \beta H \right)}\right]^2 }$
\\
 \hline
 $11s$ $\ast$
 & $\frac{2X[(1+X^2) \cosh \gamma (\phi) - 2X]}{(1-X^2)^2}$
 & $\frac{2X}{(1+X)^2}$
 & ${\tanh\left( \beta H \right)}$
\\
 \hline
 $21s$ $\ast$
 & $\frac{X^2[1-2X\cosh\gamma(\phi)+X^2]}{(1-X^2)^2}$
 & $\frac{X^2}{(1+X)^2}$
 & $\frac{\tanh\left( \beta H \right)}
         {1 + \left[ \frac{\sinh \left(2 \beta J_x \right) }
                   {\cosh \left( \beta H \right)}\right]^2 }$
\\
 \hline\hline
$10s$ $\dag$
 & $\frac{2X^2}{1-X^2}
\left\{
\frac{2\cosh^2\gamma(\phi)-1-2X\cosh\gamma(\phi)+X^2}
{1-2X\cosh\gamma(\phi)+X^2} \right.
$
 & $\frac{2X^2(1+2X)}{(1-X^2)(1+X)^2}$
 & $\frac{\tanh\left( \beta H \right)}
         {1 + \left[ \frac{\sinh \left(2 \beta J_y \right) }
                   {\cosh \left( \beta H \right)}\right]^2 }$
\\
 &$ \left.
-
\frac{X^2 [1-2X\cosh\gamma(\phi)+X^2]}{(1-X^2)^2}
\right\}$
 &
 &
\\
 \hline
$20s$ $\dag$
 & $\frac{X^4 [1-2X\cosh\gamma(\phi)+X^2]}{(1-X^2)^3}$
 & $\frac{X^4}{(1-X^2)(1+X)^2}$
 & $\frac{\tanh\left( \beta H \right)}
         {1 + \left[\frac{\sinh \left(2 \beta (J_x-J_y) \right) }
                   {\cosh \left( \beta H \right)}\right]^2 }$
\\
 \hline\hline
 \end{tabular}
 \end{center}
\label{table:class2}
\end{table}

\clearpage


\clearpage

\begin{center}
\includegraphics[angle=270,width=5in]{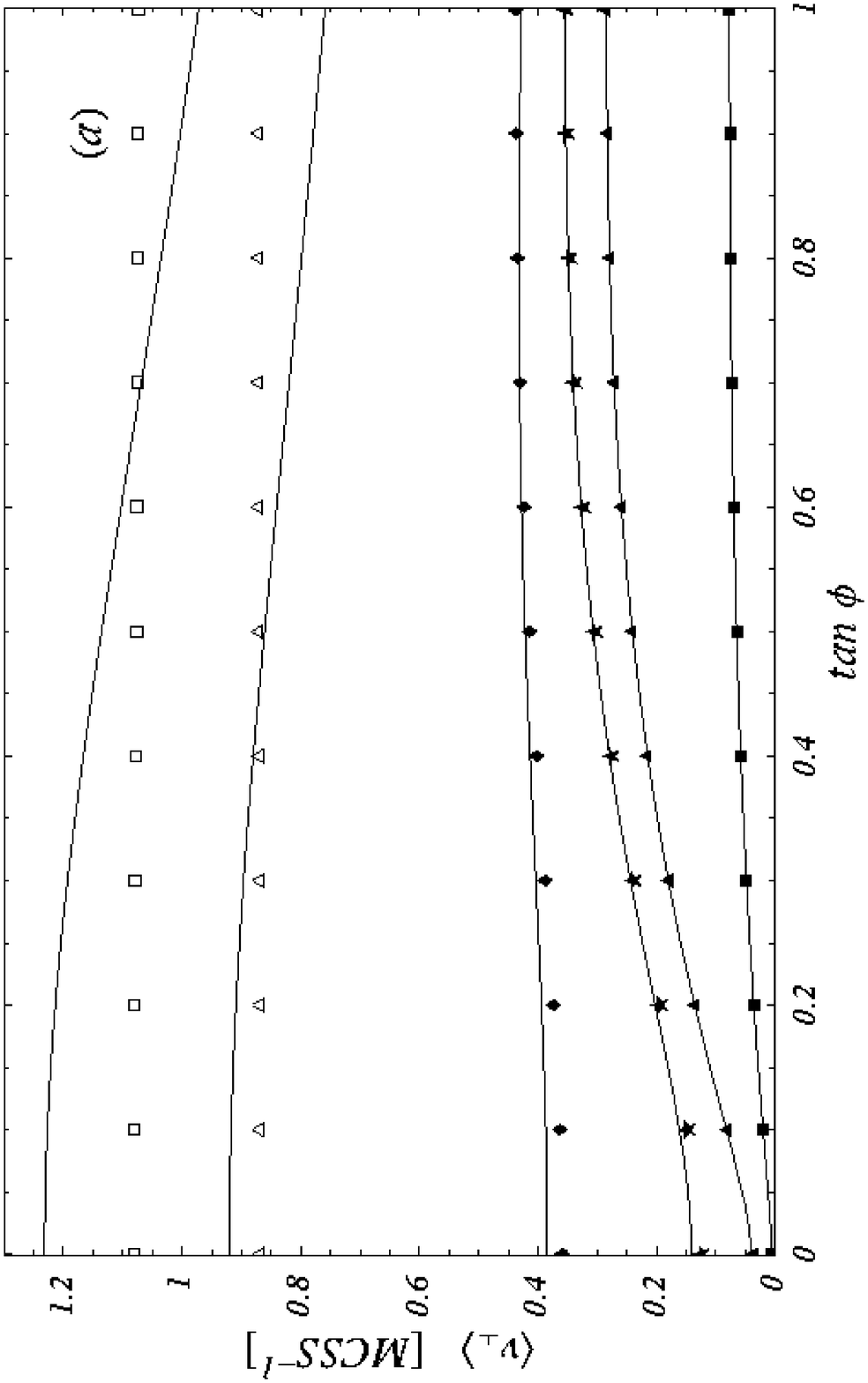}
\end{center}
\begin{center}
\includegraphics[angle=270,width=5in]{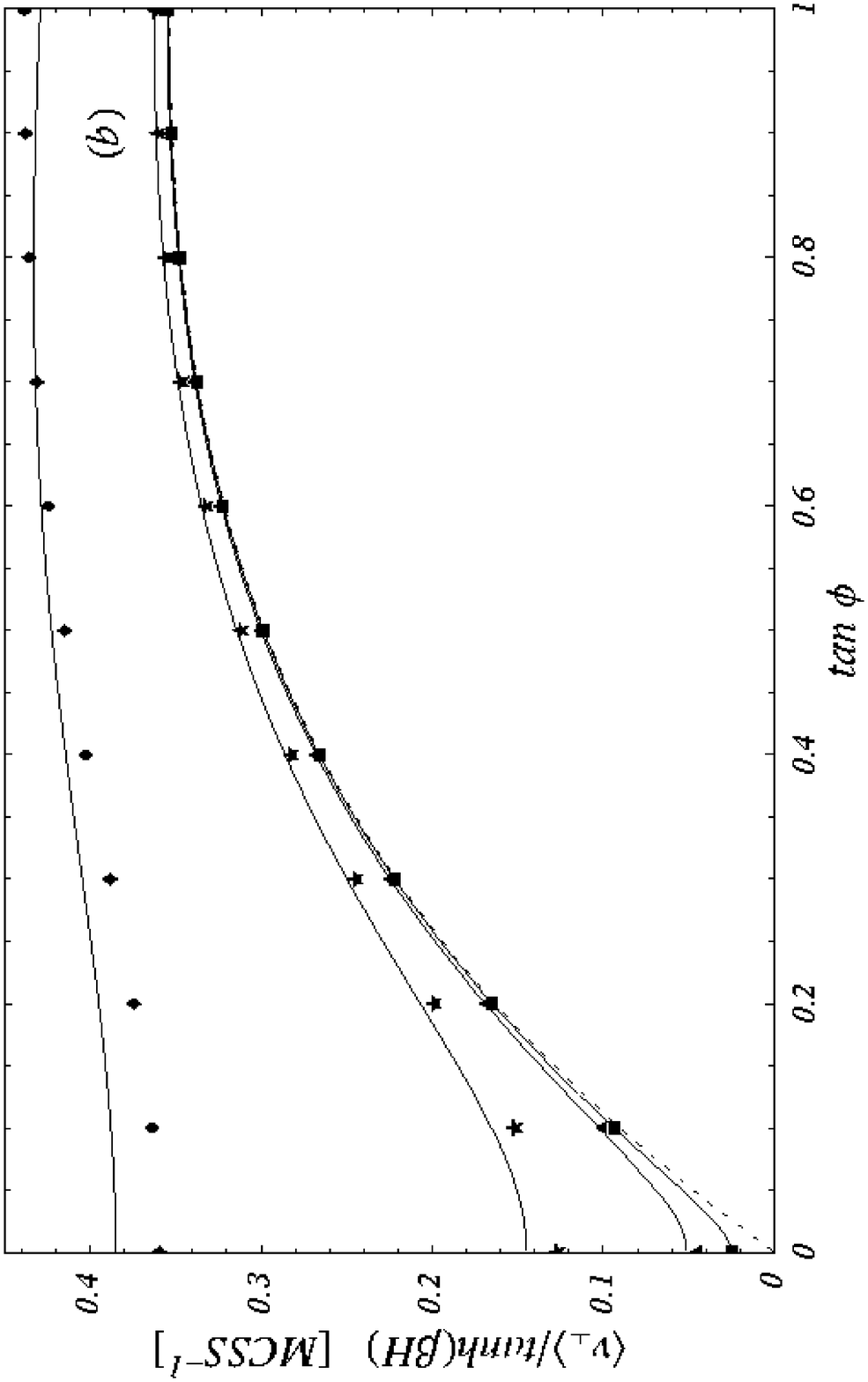}
\end{center}
\clearpage

\clearpage

\begin{center}
\includegraphics[angle=270,width=5in]{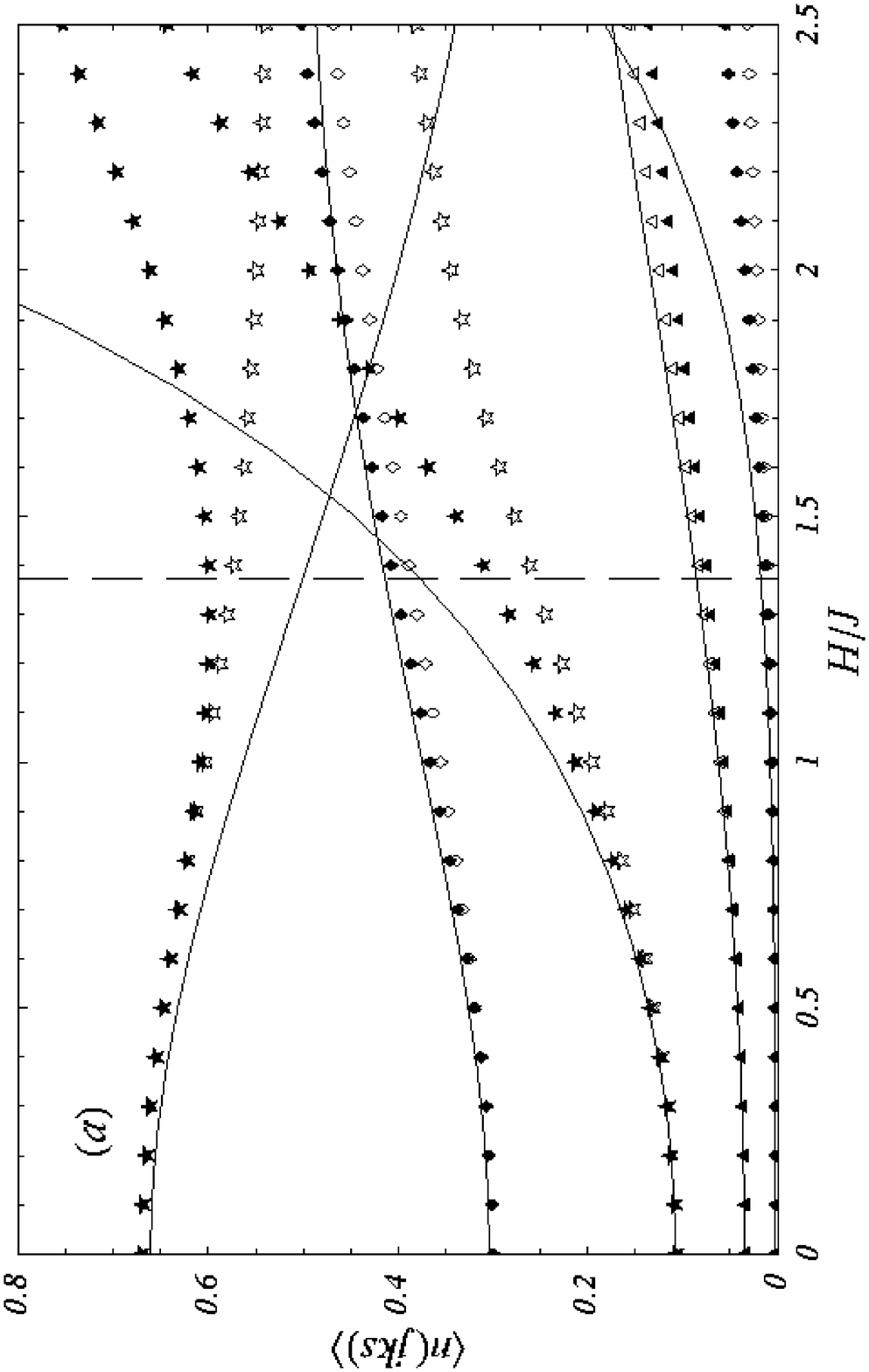}
\end{center}
\begin{center}
\includegraphics[angle=270,width=5in]{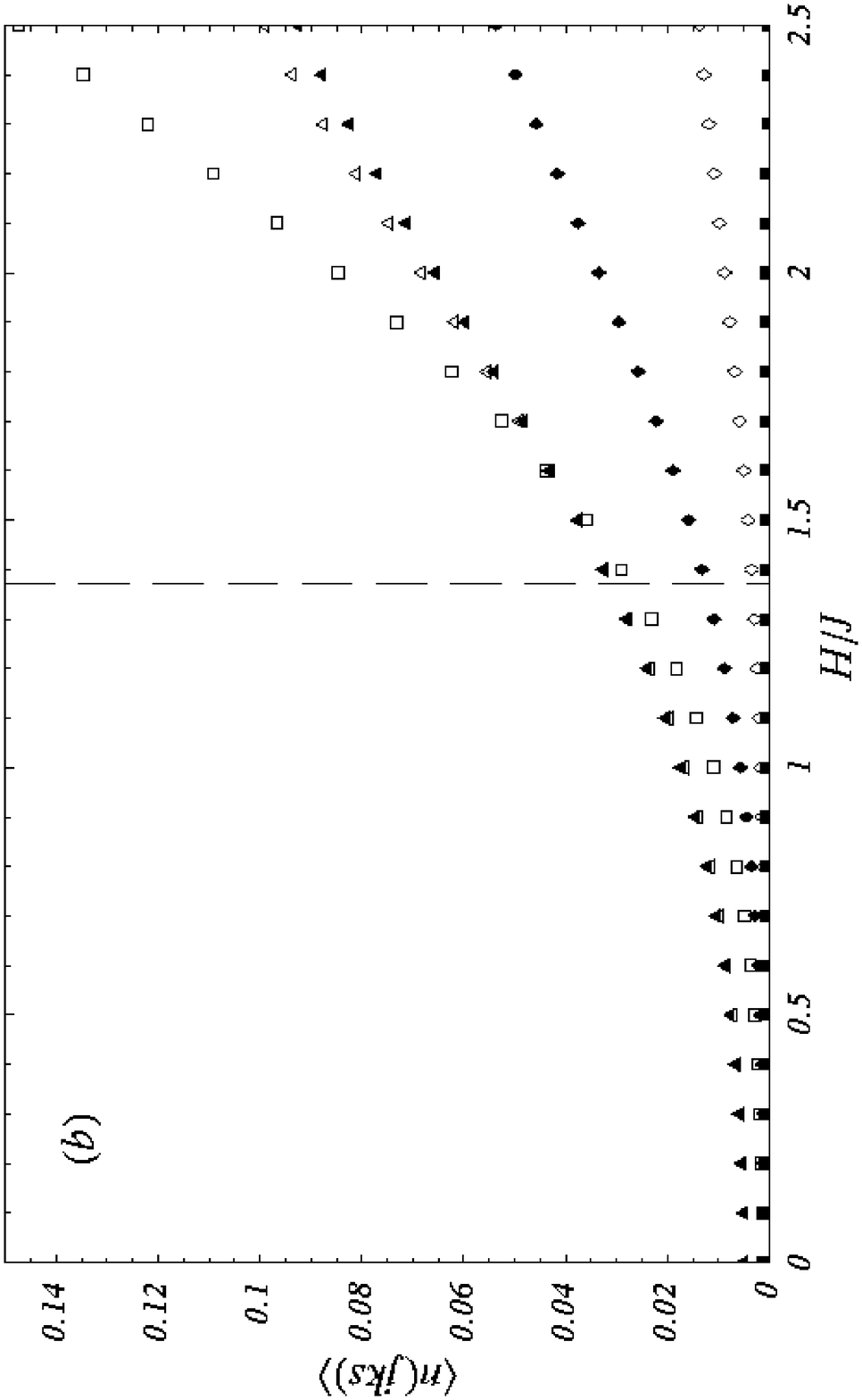}
\end{center}
\clearpage

\end{document}